\documentclass[twocolumn]{aastex62}

\usepackage{amsmath}
\usepackage{amssymb}
\usepackage{bm}

\usepackage{graphicx}
\usepackage{bold-extra}

\newcommand{\beq}{\begin{equation}}
\newcommand{\eeq}{\end{equation}}

\newcommand{\Msun}{\text{M}_\odot}

\newcommand{\Mpc}{\text{Mpc}}
\newcommand{\HI}{\text{HI}}

\newcommand{\mK}{\text{mK}}
\newcommand{\avg}[1]{\ensuremath{\langle #1 \rangle}}
\newcommand{\CIIm}{\text{C~\textsc{ii}}}
\newcommand{\OIIIm}{\text{O~\textsc{iii}}}
\newcommand{\CII}{C~\textsc{ii}}
\newcommand{\OIII}{O~\textsc{iii}}

\newcommand{\Var}[1]{\mathrm{Var}[#1]}
\newcommand{\Cov}[2]{\mathrm{Cov}[#1,#2]}

\newcommand{\upenn}{Department of Physics \& Astronomy, University of
Pennsylvania, 209 South 33rd St., Philadelphia, PA 19104, USA}
\newcommand{\cca}{Center for Computational Astrophysics, Flatiron Institute,
162 5th Ave., New York, NY 10010, USA}

\begin{document}

\title{Measuring the EoR Power Spectrum Without Measuring the EoR Power
Spectrum}

\author{Angus Beane}
\affiliation{\cca}
\affiliation{\upenn}

\author{Francisco Villaescusa-Navarro}
\affiliation{\cca}

\author{Adam Lidz}
\affiliation{\upenn}

\submitjournal{ApJ}
\shortauthors{Beane, Villaescusa-Navarro, \& Lidz}

\correspondingauthor{Angus Beane}
\email{abeane@sas.upenn.edu}

\begin{abstract}
The large-scale structure of the Universe should soon be measured at high
redshift during the Epoch of Reionization (EoR) through line-intensity
mapping. A number of ongoing and planned surveys are using the 21~cm line to
trace neutral hydrogen fluctuations in the intergalactic medium (IGM) during
the EoR. These may be fruitfully combined with separate efforts to measure
large-scale emission fluctuations from galactic lines such as [\CII], CO,
H-$\alpha$, and Ly-$\alpha$ during the same epoch. The large scale power
spectrum of each line encodes important information about reionization, with
the 21~cm power spectrum providing a relatively direct tracer of the
ionization history. Here we show that the large scale 21~cm power spectrum can
be extracted using only cross-power spectra between the 21~cm fluctuations and
each of two separate line-intensity mapping data cubes. This technique is more
robust to residual foregrounds than the usual 21~cm auto-power spectrum
measurements and so can help in verifying auto-spectrum detections. We
characterize the accuracy of this method using numerical simulations and find
that the large-scale 21~cm power spectrum can be inferred to a simulated
accuracy of within $5\%$ for most of the EoR. Our estimate of the 21~cm power
spectrum reaches $0.6\%$ accuracy on a scale of $k\sim0.1\,\Mpc^{-1}$ at
$\avg{x_i} = 0.36$ ($z=8.34$ in our model). An extension from two to $N$
additional lines would provide $N(N-1)/2$ cross-checks on the large-scale
21~cm power spectrum. This work strongly motivates redundant line-intensity
mapping surveys probing the same cosmological volumes.

\end{abstract}


\section{Introduction}\label{sec:intro}

Upcoming 21~cm surveys are poised to make a first detection of redshifted
21~cm fluctuations from the EoR within the next several years
\citep{DeBoer:2016tnn}. These measurements will provide a direct probe of the
distribution of neutral hydrogen in the IGM, revealing the spatial structure
of the reionization process, and its redshift evolution. Along with these
measurements, several other ``line-intensity'' mapping surveys are planned to
map out large-scale structure in the galaxy distribution using convenient
emission lines with current targets including [\CII], CO, Ly-$\alpha$, and
H-$\alpha$ \citep[see e.g.][and references therein]{kovetz2017:im_review}.
These surveys study the spatial fluctuations in the collective emission from
many individually unresolved sources (e.g.
\citealt{Suginohara:1998ti,Righi:2008br,Visbal10}). These measurements should
nicely complement 21~cm observations (e.g. \citealt{Lidz11,gong11:probing}):
while the 21~cm fluctuations trace-out remaining neutral hydrogen residing
mostly in the low-density IGM, the galactic emission lines track the galaxies
themselves, which presumably lie within ``bubbles'' of mostly ionized hydrogen
\citep{Lidz:2008ry}.

In fact, recent work has led to detections in various lines at low redshift
\citep{2010Natur.466..463C,2001defi.conf..241D,Keating:2016pka,2018MNRAS.478.1911P,2016MNRAS.457.3541C,Croft:2018rwv},
bolstering efforts to employ the line-intensity mapping technique at earlier
times during the EoR. It is hence timely to explore the scientific benefits of
combining 21~cm observations of the EoR with line-intensity mapping surveys in
other emission lines.

Here we consider, for the first time, one potential advantage of combining
21~cm surveys of the EoR with line-intensity mapping surveys in {\em two
additional lines.} Specifically, we show that the linear bias factor of the
21~cm field may be extracted solely from cross-power spectra between the 21~cm
fluctuations and those in each of two separate lines. This can provide an
important cross-check on inferences from the 21~cm auto-power spectrum since
cross-power spectra should be less prone to bias from residual foregrounds
\citep[e.g.][]{Furlanetto:2006pg,Lidz:2008ry}; only shared foregrounds
contribute to the average cross spectrum signal. 

The foreground problem is especially daunting in the case of redshifted
21 cm surveys, where the expected foreground-to-signal strength is on the
order of $\sim 10^5$
\citep[e.g.][]{2009A&A...500..965B,2013ApJ...768L..36P,Dillon:2013rfa}. The
basic strategy for extracting the signal is to exploit the fact that the
foregrounds should be smooth functions of frequency, while the reionization
signal has a great deal of spectral structure. In practice, this is
challenging because the instrument, calibration errors, and other effects may
imprint artificial spectral variations. Cross-spectrum measurements should be
less sensitive to such systematic effects and can therefore help confirm early
detections. For instance, \citet{2015JCAP...03..034V} show that cross-spectra
can be robustly measured even in the presence of polarized synchrotron
foregrounds; this is a troublesome case for auto-spectrum analyses because
Faraday rotation leads to frequency structure.

The amplitude of the 21~cm power spectrum evolves with redshift in a
distinctive way as reionization proceeds \citep[e.g.][]{Lidz08}, and recent
work has demonstrated that linear biasing describes the large-scale 21~cm
power spectrum rather well
\citep{McQuinn:2018zwa,Hoffmann:2018clb,2018ApJ...867...26B}. Therefore, if
our three-field method may be employed over a range of redshifts, it can be
used to extract key and robust information regarding the reionization history
of the Universe.

In recent related work we showed that the large-scale 21~cm bias factor may be
recovered using suitable cross-bispectra between the 21~cm fluctuations and
the [\CII] emission field \citep{2018ApJ...867...26B}. While the
cross-bispectra method requires only the 21~cm fluctuations and one additional
tracer field, the technique we propose here should be vastly simpler to
implement in practice (provided two additional tracers are available with
common sky and redshift coverage). This is the case because our present method
relies only on two-point statistics, and it therefore avoids practical
difficulties in carrying out cross-bispectrum analyses. For example, it is
challenging to estimate the bispectrum covariance as this involves computing a
six-point function. In addition, we will show that our present technique
allows for a more faithful extraction of the 21~cm bias factor. Ultimately,
both analyses may be carried out for additional cross-checks.

There are a broad range of possible lines that may be combined with the 21~cm
surveys. Currently, there are projects -- either ramping-up or in the planning
stages -- to perform EoR-era line-intensity surveys in: [\CII]~$158\,\mu
\text{m}$ \citep{Crites14,Lagache:2018hmk,Vavagiakis:2018gen}, rotational
transitions from CO molecules \citep{Chung:2017uot}, Ly-$\alpha$
\citep{Dore16}, and H-$\alpha$ \citep{Cooray:2016hro}. Additional
fine-structure lines such as [\OIII]~$88\,\mu \text{m}$ \citep{Moriwaki18} and
[N \textsc{ii}]~$122\,\mu \text{m}$ \citep{Serra:2016jzs} may also be suitable
--- in some cases, these lines will land in the proposed frequency bands of
the planned [\CII] surveys. The [\OIII]~$88\,\mu \text{m}$ line appears
especially promising since targeted ALMA observations around $z \sim 7-9$
galaxies have found that this line is {\em brighter} at high redshift than
expected based on local correlations between line-luminosity and
star-formation rate \citep[e.g.][and references therein]{Moriwaki18}.

In principle, one could extract the 21~cm bias using the cross-spectrum with a
traditional galaxy survey, in which case the galaxy bias may be measured
robustly from the auto-power spectrum. In practice, this is extremely
challenging because one needs {\em spectroscopic redshifts} for the galaxy
survey over a huge sky area at $z \sim 8$. If only photometric redshifts are
available, then one only accesses long-wavelength line-of-sight modes (with
small or vanishing line-of-sight wavenumbers) in the galaxy survey but
precisely these modes are lost to foreground cleaning/avoidance in the 21~cm
surveys (e.g. \citealt{Lidz:2008ry}). Fortunately, multi-line intensity
mapping provides a promising way forward here and our approach avoids
measuring bias factors from auto-spectra.

In Section~\ref{sec:approach}, we describe our three cross-spectra approach in
detail. In Section~\ref{sec:simulations} we briefly discuss the radiative
transfer simulations of reionization \citep{2007MNRAS.377.1043M,Lidz08} used
in our analysis, the reionization model assumed, and our method for generating
mock line-intensity mapping data cubes. We then quantify the accuracy of our
technique in Section~\ref{sec:results}. The survey specifications required to
extract bias factors with this method are discussed briefly in
Section~\ref{sec:detectability}. We conclude in Section~\ref{sec:conclusions}.
We assume a $\Lambda$CDM cosmology, parameterized by $(\Omega_m,
\Omega_{\Lambda}, \Omega_b, h, \sigma_8, n_s) = (0.27, 0.73, 0.046, 0.7, 0.8,
1)$ as in the simulations used in this work \citep{McQuinn:2007dy}. While
these parameters differ slightly from presently favored values (e.g.
\citealt{2018arXiv180706209P}), this should not impact our conclusions.

\section{Approach}\label{sec:approach}
Here we define terms and describe our three cross-spectra approach. Ignoring
redshift-space distortions and spin-temperature fluctuations, the 21~cm
brightness temperature contrast between neutral hydrogen gas and the cosmic
microwave background is:
\beq\label{eq:brightness_temp}
T_{21}(\bm{x}) = T_0 X_{\HI}(\bm{x})[1+\delta_\rho(\bm{x})]\text{.}
\eeq
Here $T_0 = 28\,\mK[(1+z)/10]^{1/2}$ \citep[e.g.][]{Zaldarriaga:2003du},
$X_{\HI}(\bm{x})$ is the neutral hydrogen fraction at position $\bm{x}$, and
$\delta_\rho(\bm{x})$ is the gas density contrast, which is assumed to follow
the overall matter density field on the large scales of interest. Although
ionized regions imprint large-scale fluctuations in the 21~cm field, on scales
much larger than the size of the ionized regions, the 21~cm fluctuations
should nevertheless follow a linear biasing relation
\beq\label{eq:21cm_bias}
T_{21}(\bm{k}) = \pm \avg{T_{21}} b_{21} \delta_{\text{lin}}(\bm{k})\text{,}
\eeq
where the $\pm$ indicates that the fields are either correlated ($+$) or
anti-correlated ($-$) --- during the bulk of the EoR, the 21~cm and density
fields are anti-correlated on large scales in most models
\citep[e.g.][]{Lidz:2008ry}. Here $T_{21}(\bm{k})$ is the Fourier transform of
the brightness temperature field (Equation~\ref{eq:brightness_temp}) and
$\delta_{\text{lin}}(\bm{k})$ is the Fourier transform of the linear density
contrast.\footnote{Our Fourier convention is: $T_{21}(\bm{k}) = \int
\text{d}^3x\, T_{21}(\bm{x}) e^{i \bm{k} \cdot \bm{x}}$ and $T_{21}(\bm{x}) =
\int \frac{\text{d}^3k}{(2\pi)^3}\, T_{21}(\bm{k}) e^{-i \bm{k} \cdot
\bm{x}}$.} The quantity $b_{21}$ is the dimensionless, and scale-independent,
linear bias factor of the 21~cm fluctuation contrast, $\delta_{21}(\bm{x}) =
\left(T_{21}(\bm{x}) - \avg{T_{21}}\right)/\avg{T_{21}}$, while the
$\avg{T_{21}}$ factor reverts to brightness temperature units (since the
average brightness temperature is not itself observable from interferometric
measurements.) In this work when we refer to the ``bias'' we mean
$\avg{T_{21}}b_{21}$ (and likewise for the intensity mapping surveys.)

Likewise, we can consider additional tracer lines, such as [\CII]. On large
scales, the Fourier transform of the specific intensity of each of these lines
should be well-described by
\beq\label{eq:linear_biasing}
I_{i}(\bm{k}) = \avg{I_{i}} b_{i}  \delta_{\text{lin}}(\bm{k})\text{,}
\eeq
where $\avg{I_{i}}$ is the mean specific intensity of the emission
line.\footnote{We follow standard conventions in expressing 21~cm fluctuations
in brightness temperature units, i.e. in $\mK$, while we use specific
intensity units for the other tracer lines, i.e. $I_{i}$ is the specific
intensity in $\text{Jy/str}$.} For the case of emission lines sourced by gas
within galaxies, the relevant bias factor is the luminosity-weighted bias of
the line-emitting host halos (e.g. \citealt{Lidz11}). To be completely general
we should also include a $\pm$ here (as in Equation~\ref{eq:21cm_bias}), but
for the galactic emission lines we generally expect brighter line emission in
overdense regions.

On sufficiently large scales, the auto-power spectrum of the fluctuations in
each tracer line (Equation~\ref{eq:linear_biasing}) will be
\beq\label{eq:bias_ps}
\begin{split}
P_{i, i}(k, z) &\equiv \avg{I_{i}(k, z) I_{i}^{*}(k, z)} \\
&= \left[\avg{I_{i}}(z) b_{i}(z)\right]^2 P_{\text{lin}}(k, z)\text{,}
\end{split}
\eeq
where $P_{\text{lin}}(k,z)$ is the linear matter power spectrum. Similarly, on
large scales the 21~cm auto-power spectrum should follow $P_{21,21}(k,z) =
\left[\avg{T_{21}}(z) b_{21}(z)\right]^2 P_{\text{lin}}(k,z)$. In principle,
one can infer the bias factors $\avg{I_i} b_i$ and $\avg{T_{21}} b_{21}$ from
auto-power spectrum measurements (assuming a model for the linear power
spectrum). However, foreground cleaning/avoidance present significant
challenges here
\citep[e.g.][]{2012MNRAS.419.3491L,2013ApJ...769..154M,2015ApJ...804...14T,2016ApJ...819....8P,Ewall-Wice:2016bhu}
and residual foregrounds may bias such inferences.

Another approach is to measure the cross-power spectrum between two lines $i$
and $j$. In this case, one measures
\beq\label{eq:xps}
P_{i,j} = r_{i, j} \avg{I_{i}}  \avg{I_{j}} b_{i} b_{j} P_{\text{lin}}\text{,}
\eeq
where $r_{i,j}$ is the cross-correlation coefficient which ranges from $-1$ to
$1$.\footnote{Note that here we adopt the convention that the bias factors are
always positive and that the sign of the cross-spectrum is determined solely
by that of the correlation coefficient. This convention differs from our
previous work \citep{2018ApJ...867...26B}.} In the above equation and in what
follows, we generally suppress redshift and wavenumber labels for brevity. In
general, $r_{i,j}$ is scale-dependent, but asymptotes to $-1$ (for
anticorrelated fields) or $1$ (for correlated fields) on large
scales.\footnote{Note that we neglect shot-noise contributions to the
auto-spectrum in Equation~\ref{eq:bias_ps}, as well as correlated shot-noise
terms in the cross-power spectrum. This should be a very good approximation on
the scales of interest unless the line-emitting sources are quite rare (e.g.
\citealt{lidz2016:remove_interloper}). Even in the case of rare sources, the
shot-noise term should be a white-noise contribution on scales much larger
than the size of the host halos. In this case, one can perform a joint fit for
the shot-noise along with the clustering terms.} If one of the lines is the
21~cm field, we replace $\avg{I_i}$ with $\avg{T_{21}}$ in
Equation~\ref{eq:xps}.

However, in the presence of a third line $k$, and with $P_{j,k}$ and $P_{k,i}$
defined analogously as in Equation~\ref{eq:xps}, we can simply write
\beq\label{eq:threefields}
\begin{split}
P_{i,i}=(\avg{I_i} b_i)^2 P_{\text{lin}} &= \frac{r_{j,k}}{r_{i,j} r_{k,i}} \frac{P_{i,j} P_{k, i}}{P_{j,k}} \\
&\equiv R_{i,j,k} P_{i,j,k}\text{,}
\end{split}
\eeq
where we have defined $R_{i,j,k} \equiv r_{j,k}/(r_{i,j} r_{k,i})$ and
$P_{i,j,k} \equiv (P_{i,j}P_{k,i})/P_{j,k}$. On sufficiently large scales,
$R_{i,j,k} \rightarrow 1$, but on intermediate scales $R_{i,j,k} > 1$ for most
reasonable cases when the various $r$'s are close in magnitude.
Equation~\ref{eq:threefields} shows that (on sufficiently large scales where
linear biasing holds and $R_{i,j,k} \sim 1$) we can recover the linear bias
factor of field $i$ from a suitable ratio of cross-spectra. Here we suppose
that the underlying density power spectrum is well known.
Equation~\ref{eq:threefields} is the main point of this paper; in the
remainder of this work we consider an application to the EoR and quantify its
accuracy. Specifically, we will test the range of validity -- in spatial scale
and redshift/ionization fraction -- of the assumption that $R_{i,j,k}=1$,
along with the linear biasing approximations of
Equations~\ref{eq:21cm_bias}~\&~\ref{eq:linear_biasing}. Note that testing the
assumption that $R_{i,j,k} = 1$ directly from upcoming data will require
reliable auto-spectra.

We turn now to the specific case of EoR surveys with the goal of extracting
the 21~cm bias factor using only cross-power spectra. For further specificity
we suppose that the two additional tracer lines are [\CII] and [\OIII],
although little of the analysis that follows depends on the choice of these
two lines --- any of the lines mentioned in Section~\ref{sec:intro} can be
used instead of [\CII] or [\OIII]. In this case,
Equation~\ref{eq:threefields} may be applied as
\beq\label{eq:threefields_specific}
\begin{split}
P_{21,21} &= (\avg{T_{21}} b_{21})^2 P_{\text{lin}}\\
&= \frac{P_{21,\CIIm} P_{\OIIIm, 21}}{P_{\CIIm, \OIIIm}}\text{,}
\end{split}
\eeq
i.e. assuming $R_{21,\CIIm,\OIIIm} = 1$.

We expect this approach to break down on small scales. First, the three fields
will be well-correlated (or anti-correlated) only on large scales, with the
21~cm field and the [\CII], [\OIII] fields decorrelating on scales smaller
than the size of the ionized regions \citep{Lidz11}. Second, we assume linear
biasing which should break down on scales where second-order bias terms become
significant \citep{McQuinn:2018zwa}.

One caveat here is that we neglect redshift space distortions throughout.
Including these effects will make the power spectra in
Equation~\ref{eq:threefields_specific} angle-dependent. Although these effects
are well studied in the case of the 21~cm auto-spectrum (e.g.
\citealt{Mao12}), an extension of our three cross-spectra method may be needed
to account for these distortions.

\section{Simulations}\label{sec:simulations}

In order to investigate the accuracy of Equation~\ref{eq:threefields_specific}
we turn to $(186\,\Mpc)^3$ radiative transfer simulations of the EoR
\citep{2007MNRAS.377.1043M,McQuinn:2007dy,Lidz08}. In these calculations,
radiative transfer is post-processed onto a $(1024)^3$ dark matter only
simulation run with \texttt{GADGET-2} \citep{Springel:2005mi}. The dark matter
simulation resolves halos only down to $10^{10}\,\Msun$, however halos down to
$10^8\,\Msun$ are added manually in post-processing with the correct
statistical properties \citep{2007MNRAS.377.1043M}. Halos resolved directly in
the simulation (i.e. $>10^{10}\,\Msun$) are identified with a
Friends-of-Friends algorithm with a linking length of 0.2.
 
 In what follows, we adopt the abundant mini-halo sink scenario
 \citep{2007MNRAS.377.1043M,Lidz08} as our baseline reionization model.
 Although the detailed model for photon sinks implemented in these simulations
 may not be fully realistic, the smaller ionized regions in ``abundant sink''
 scenarios may, in fact, be more plausible than the other cases considered in
 this previous work \citep{McQuinn:2018zwa}. In any case, the accuracy of our
 method does not depend strongly on the precise reionization model assumed.

In order to model the [\CII] and [\OIII] emission fluctuations, we assume that
the luminosity in each line is correlated with the host halo mass.
Specifically, we adopt a power-law average relation between line-luminosity
and halo mass:
\beq\label{eq:im_form}
\avg{L_i}(M) = L_{i,0} \left[\frac{M}{M_0}\right]^{\alpha_i},
\eeq
where $M$ is the mass of the halo, $\avg{L_i}$ is the average luminosity, and
$L_{i,0}$ is the luminosity at characteristic mass $M_0$. In order to account
for scatter in this relation, we add a random number so that each halo's
luminosity is $L_i = \avg{L_i}(1 + \epsilon)$ where $\epsilon$ is drawn from a
zero-mean lognormal distribution of width 0.4 dex.

In what follows we assume that each host halo in the simulation hosts a [\CII]
and [\OIII] emitter. If only a random fraction $f$ of halos host active [\CII]
and/or [\OIII] emitters while $L_{i,0}$ is boosted to fix the average
specific-intensity in each line, this does not change the 21~cm-[\CII] or
21~cm-[\OIII] cross-power spectra. This represents the case that
star-formation activity has a short duty-cycle, yet the total star-formation
rate density is fixed to the observed value. If the same random fraction emit
in both [\CII] and [\OIII] this can boost the cross-shot noise contribution to
$P_{\CIIm,\OIIIm}$, but this is highly sub-dominant on the scales of interest
($k \leq 0.4$ Mpc$^{-1}$) even for $f=10^{-2}$.

In order to estimate the specific intensity of the two fields, we use nearest
grid-point interpolation to estimate the emissivity on a $512^3$ Cartesian
grid, matching the resolution of the density and 21~cm fields from
\citet{Lidz08}. Note that we can test the accuracy of
Equation~\ref{eq:threefields_specific} without specifying the numerical value
of $L_{i,0}$ or $M_0$ since they cancel in the ratio. The value of $\alpha_i$,
on the other hand, controls which host-halos (and galactic star-formation
rates) produce most of the specific intensity in line $i$.\footnote{Note we
assume that the minimum host halo mass of the [\CII] and [\OIII] emitters is
$10^8 M_\odot$, comparable to the atomic cooling mass.  The true minimum host
mass of the emitters may, in fact, be larger. However, note that the average
specific intensity may be fixed by the total star-formation rate density and
the line-luminosity star-formation rate correlation. Provided these quantities
are fixed, then the main impact of boosting the minimum host halo mass will be
to increase slightly the bias factors, $b_i$, and the signal strength. See
e.g. \citep{lidz2016:remove_interloper} for more details regarding
line-intensity fluctuation models.} If the value of $\alpha_i$ is the same for
[\CII] and [\OIII], then the two fields differ only by an overall
multiplicative factor and Equation~\ref{eq:threefields_specific} reduces to a
simple ratio between a single cross-spectrum and an
auto-spectrum.\footnote{This assumes, as we do here, that the scatter in the
luminosity-mass relation is perfectly correlated between [\CII] and [\OIII] at
fixed $\alpha_i$.}

We consider three different values for $\alpha_i$: $2/3$, $1$, and $4/3$. We
refer to these as L, M, and H since they provide most weight to low, medium,
and high mass host-halos respectively. We allow for the case that the two
lines have different values of $\alpha_i$: i.e., we consider 21~cm-L-M,
21~cm-M-H, and 21~cm-H-L, with L, M, or H standing in for [\CII] or [\OIII] in
Equation~\ref{eq:threefields_specific}. We then measure the various
cross-spectra using a slightly modified version of the power spectrum
calculator in \texttt{21cmFAST} \citep{Mesinger11,2018arXiv180908995P}.

\section{Results}\label{sec:results}

We first investigate how well our three cross-spectra approach for measuring
the large-scale 21~cm bias agrees with the true bias. We measure the true bias
as
\beq\label{eq:truebias}
\avg{T_{21}} b_{21}(k) \equiv \sqrt{\frac{P_{21,21}(k)}{P_{\delta,\delta}(k)}}\text{,}
\eeq
and also estimate the bias as
\beq\label{eq:truebias_cross}
\avg{T_{21}} b_{21}(k) \simeq
\left|\frac{P_{21,\delta}(k)}{P_{\delta,\delta}(k)} \right| \text{,}
\eeq
where $P_{\delta,\delta}(k)$ is the auto-power spectrum of the simulated
density field and $P_{21,\delta}(k)$ is the 21~cm-density cross-power
spectrum. Note that Equation~\ref{eq:truebias_cross} assumes that the
correlation coefficient $\left|r_{21,\delta}\right| = 1$ and so will depart
from Equation \ref{eq:truebias} on small scales, but the two should converge
on large scales (see Section~\ref{sec:intro}). The absolute value in
Equation~\ref{eq:truebias_cross} comes about from the convention adopted in
Section~\ref{sec:approach}. On large scales where the 21~cm, [\CII], and
[\OIII] fields are each well correlated or anti-correlated with the density
field and linear theory applies, we expect all estimates of $\avg{T_{21}}
b_{21}$ to agree. When we estimate the bias factors using our three
cross-spectra method (Equation~\ref{eq:threefields_specific}) we use the
simulated density power-spectrum, since this is extremely close to the linear
theory prediction on the relevant scales and redshifts.

\begin{figure}
\includegraphics[width=\columnwidth]{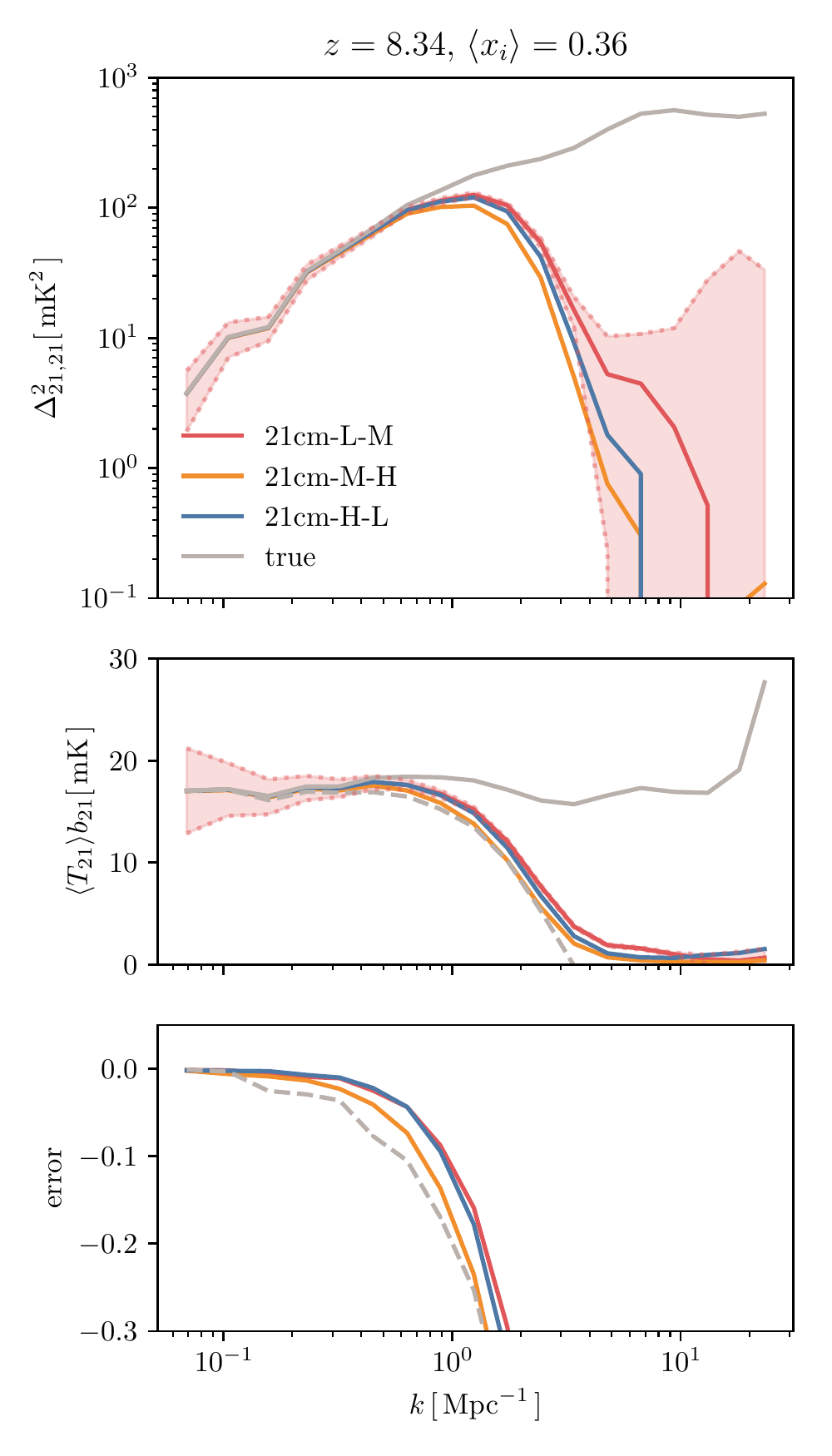}
\caption{{\em Upper:} The simulated, dimensionless 21~cm auto-power spectrum
(gray) compared to that inferred from our three-cross spectra approach
assuming linear biasing at $\avg{x_i}=0.36$, $z=8.34$. The different colors
correspond to various possible line-luminosity mass relations (L, M, H), as
described in Section~\ref{sec:simulations}. The shaded area shows the
$1\,\sigma$ expected errors for the 21~cm-L-M survey described in
Section~\ref{sec:detectability}. {\em Middle:} The 21~cm bias factor extracted
from our three cross-spectra approach in the different line-luminosity models.
These are compared with that inferred from the 21~cm auto-spectrum (solid
gray) and the 21~cm-density cross-spectrum (gray dashed). {\em Bottom:} The
relative difference between the different bias-factor models. On large scales
all inferences agree.}
\label{fig:b21_vs_k}
\end{figure}

The bias factors inferred from Equation~\ref{eq:threefields_specific} are
shown in Figure~\ref{fig:b21_vs_k} for each of the three combinations of our
luminosity-mass relation models (L-M, M-H, H-L) at $z=8.34$ when the model
volume-averaged ionization fraction is $\avg{x_i}=0.36$. These are compared
with the bias inferred from the 21~cm auto-spectrum
(Equation~\ref{eq:truebias}) and the 21~cm-density cross-spectrum
(Equation~\ref{eq:truebias_cross}). On large scales ($k \lesssim
0.3\,\Mpc^{-1}$), the methods converge to very nearly the same value. We find
that on a scale of $k=0.1\,\Mpc^{-1}$ at $\avg{x_i}=0.36$ the three methods
agree with the true value to within $0.6\%$. In the case of 21~cm-L-L,
21~cm-M-M, or 21~cm-H-H models the agreement is slightly worse but still at
the percent-level. Note that another approach for estimating the 21~cm bias
would use only the 21~cm-[\CII] cross-spectrum and the [\CII] auto-spectrum.
This requires measuring the [\CII] auto-spectrum, which is subject to
contamination from interloping line emission, and so we pursue only the more
robust three-field technique here.

The success results because the ionized regions are sufficiently smaller than
this scale ($k=0.1\,\Mpc^{-1}$), ensuring that the 21~cm and line-intensity
fields are highly anti-correlated and that second-order biasing contributions
are small. For example, the cross-correlation coefficient between the 21~cm
field and the density field is $r_{21,\delta} = -0.99$ at $k=0.1\,\Mpc^{-1}$
for $\avg{x_i}=0.36$, $z=8.34$.

On smaller scales, our approach breaks down. At $\avg{x_i}=0.36$, the
different bias factor estimates begin diverging at the $\geq 10\%$ level near
$k \sim 0.4\,\Mpc^{-1}$. This occurs because the fields start to de-correlate
and second order biasing terms become more important. As anticipated after
Equation~\ref{eq:threefields}, the three cross-spectra approach underestimates
the bias factor in this regime. This underestimation may allow one to place
robust lower limits on $P_{21,21}$ that are only $\sim50\%$ smaller than the
true value down to $k\sim2\,\Mpc^{-1}$ at this stage of the EoR, although the
model-dependence of such limits warrants further investigation.

\begin{figure}
\includegraphics[width=\columnwidth]{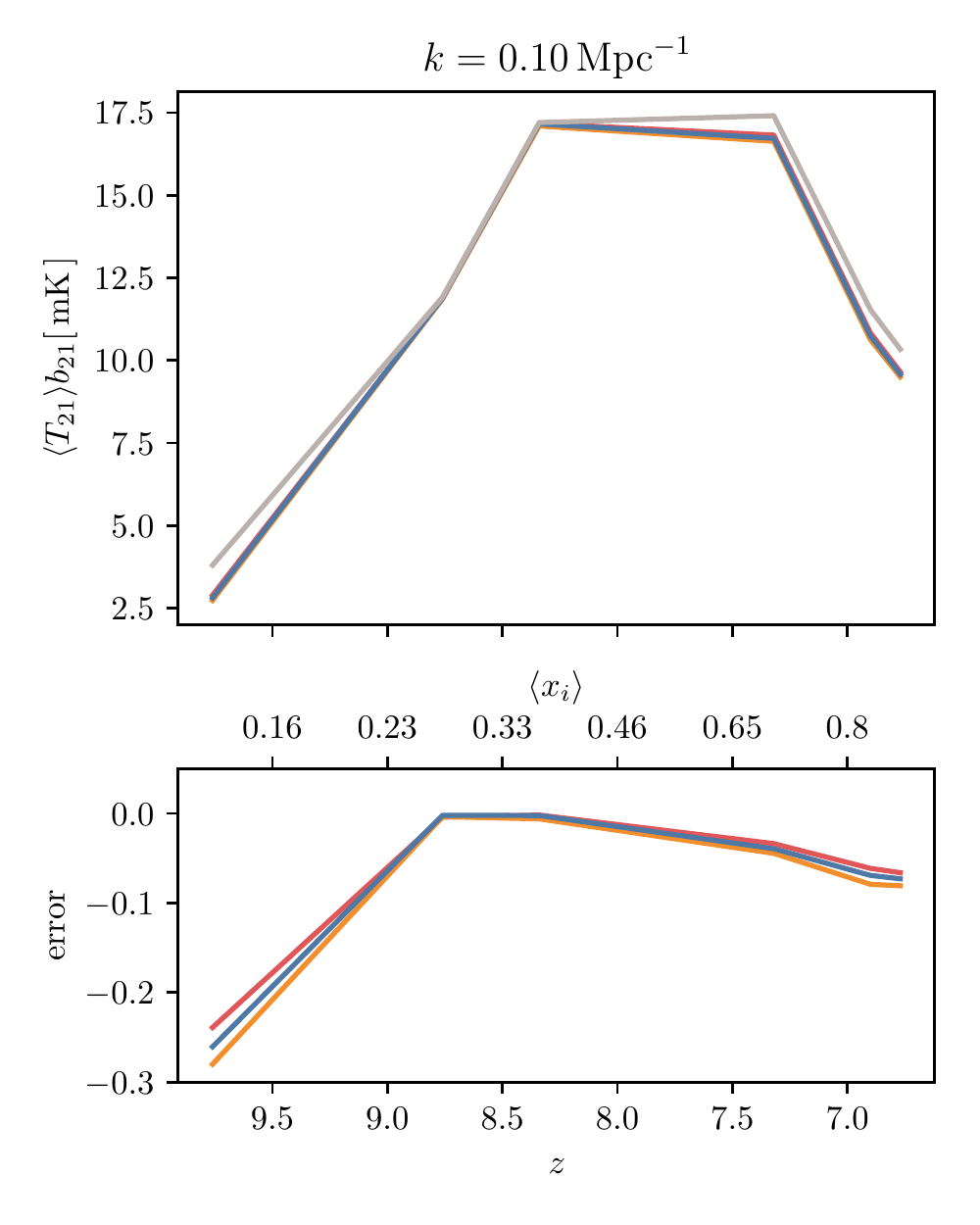}
\caption{{\em Upper:} The inferred 21~cm bias factor as a function of
redshift/volume-average ionization fraction at $k=0.1\,\Mpc^{-1}$. The
different colored lines show inferences from our three cross-spectra approach
in the different line-luminosity models (see Figure~\ref{fig:b21_vs_k} and
text). The gray line shows the true bias factor measured from the
21~cm-$\delta$ cross-spectrum. {\em Lower:} The relative error in the three
cross-spectra approach. We find better than $5\%$ agreement for most of the
EoR with sub-percent accuracy achieved near $\avg{x_i}=0.36$ at $z=8.34$. At
$\avg{x_i} \sim 0.15$ the fields decorrelate on large scales and so the
approach breaks down (see text).}
\label{fig:b21_vs_z}
\end{figure}

At later times, the average ionization fraction and the bubble sizes increase
and so the scale at which the linear biasing approximation breaks down moves
to larger scales. For example, at $\avg{x_i} = 0.7$ ($z = 7.32$), the approach
breaks down at the $\sim 10\%$ level at $k\sim0.3\,\Mpc^{-1}$, though an
accuracy of only a few percent is achieved at the largest scales considered
here. We suspect better agreement on even larger scales than probed by our
relatively small simulation volume.

In Figure~\ref{fig:b21_vs_z} we turn to consider the redshift evolution of the
21 cm bias factor at $k=0.1\,\Mpc^{-1}$. As emphasized earlier, the redshift
evolution of the 21~cm bias factor encodes interesting information about how
reionization proceeds. The three cross-spectra method generally recovers the
overall evolution of the 21~cm bias factor with redshift and volume-averaged
ionization fraction quite accurately. This suggests that our technique may
help in reconstructing the reionization history of the Universe, or in
verifying the results from 21~cm auto-spectrum measurements.

The one exception is near $\avg{x_i} \sim 0.15$, where our technique is
relatively inaccurate. This occurs because large-scale overdense regions are
initially brighter in 21~cm than typical regions in our model and so the 21~cm
fields are intially {\em positively-correlated} with the density fluctuations.
As reionization begins, the large-scale overdense regions ionize first which
causes the correlation coefficient between the 21~cm and density fields to
reverse signs. Consequently, there is an intermediate period (near $\avg{x_i}
\sim 0.15$ in this model) where the two fields are roughly {\em uncorrelated}
on large scales \citep{Lidz08}. This causes our method to break down, although
we caution that incorporating spin-temperature fluctuations into the modeling
may modify this conclusion. Note also that it will be challenging to perform
line-intensity mapping observations at very early times before, e.g.,
sufficient metal enrichment occurs.

While our baseline model assumes the abundant mini-halo sinks scenario we have
also investigated the fiducial model used in \citet{Lidz08}. Although this
latter model has a different ionization history and bias factor evolution, the
accuracy of our three cross-spectra method is broadly similar in this case.
For example, near the midpoint of reionization in this model ($z=7.32,
\avg{x_i}=0.54$), the 21~cm bias extraction also reaches sub-percent accuracy.

\section{Detectability}\label{sec:detectability}

Encouraged by the success of our approach in simulations, we briefly describe
the survey specifications required to infer 21~cm bias factors using this
technique. Here we consider only rough estimates and defer an in depth
treatment of noise power spectra, variance from residual foregrounds, and a
full probabilistic, multi-field framework to future work.

We first describe the relevant variance and covariance formulae (for
derivations, see e.g. \citealt{2015JCAP...03..034V}):
\beq\label{eq:var_covar}
\begin{split}
\Var{P_{i,j}} &= P_{i,j}^2 + P_{i,\text{tot}}P_{j,\text{tot}} \\
\Cov{P_{i,j}}{P_{i,k}} &= P_{i,\text{tot}}P_{j,k} + P_{i,j}P_{i,k}\text{,}
\end{split}
\eeq
where $P_{i,\text{tot}} = P_{i} + N_{i}$ and $N_{i}$ is the instrumental noise
power spectrum of line $i$. For simplicity, we neglect the shot-noise
contribution to each field. We note that Equation~\ref{eq:var_covar} is only
valid in the Gaussian approximation, but this is suitable for the large scales
of interest in our approach.

We can now apply the standard propagation of errors formula to
Equation~\ref{eq:threefields_specific} and substitute
Equation~\ref{eq:var_covar}, yielding:
\beq\label{eq:noise_P21}
\begin{split}
&\Var{P_{21}} = \\
& \left(\frac{P_{21,\CIIm}}{P_{21,\OIIIm}}\right)^2\left(P_{21,\OIIIm}^2 + P_{21,\text{tot}}P_{\OIIIm,\text{tot}}\right) \\
&+ \left(\frac{P_{21,\OIIIm}}{P_{21,\CIIm}}\right)^2\left(P_{21,\CIIm}^2 + P_{21,\text{tot}}P_{\CIIm,\text{tot}}\right) \\
&+ \left(\frac{P_{21,\CIIm}P_{21,\OIIIm}}{P_{\CIIm,\OIIIm}^2}\right)^2\left(P_{\CIIm,\OIIIm}^2 + P_{\CIIm,\text{tot}}P_{\OIIIm,\text{tot}}\right) \\
&+ \frac{P_{21,\CIIm}P_{21,\OIIIm}}{P_{\CIIm,\OIIIm}^2}\left(P_{21,\text{tot}}P_{\CIIm,\OIIIm} + P_{21,\CIIm}P_{21,\OIIIm} \right) \\
&- \frac{P_{21,\CIIm}^2P_{21,\OIIIm}}{P_{\CIIm,\OIIIm}^3}\left(P_{\OIIIm,\text{tot}}P_{21,\CIIm} + P_{21,\OIIIm}P_{\CIIm,\OIIIm} \right) \\
&- \frac{P_{21,\CIIm}P_{21,\OIIIm}^2}{P_{\CIIm,\OIIIm}^3}\left(P_{\CIIm,\text{tot}}P_{21,\OIIIm} + P_{21,\CIIm}P_{\CIIm,\OIIIm} \right)\text{.}
\end{split}
\eeq

The number of modes in a bin of width $\delta k$ centered on $k$ is,
\beq\label{eq:num_modes}
N_m = \frac{4\pi k^2 \delta k}{V_{\text{fund}}}\text{,}
\eeq
where $V_{\text{fund}}$ is the volume of a fundamental mode. We assume a
square survey area and therefore compute,
\beq\label{eq:vfund}
V_{\text{fund}} = \frac{(2\pi)^3}{L_{\bot}^2L_{\parallel}}\text{,}
\eeq
where $L_{\bot}$ is the side length of the survey area and $L_{\parallel}$ is
the length of the redshift bin $\Delta z$.

We assume a joint survey area of $100\,\text{deg}^2$ and bin widths of $\delta
k = 0.03 \Mpc^{-1}$ and $\Delta z = 0.25$. In order to make a rough estimate,
we assume that each experiment reaches sample-variance limited sensitivity at
$k=0.1\,\Mpc^{-1}$, with $N_i=P_i$ at this wavenumber and adopt a pure,
isotropic white-noise power spectrum. In the case of [\CII], the required
noise depends on the uncertain average specific intensity which determines, in
part, the signal strength, $P_i$. A plausible value is $\avg{I_{\CIIm}}=5
\times 10^2\,\text{Jy/str}$ at $z=8.34$ \citep{2018ApJ...867...26B}. In this
case, $N_{\CIIm} = 1.6 \times 10^9$, $2.5 \times 10^9$, $3.9 \times
10^9\,(\text{Jy}/\text{str})^2\,\Mpc^3$ for the L, M, and H models of the
[\CII] line, respectively at $z=8.34$. These noise requirements are comparable
to the values forecasted for Stage-II [\CII] line-intensity mapping forecasts
in \citet{silva15:prospects,lidz2016:remove_interloper}. We expect broadly
similar noise requirements for hypothetical future [\OIII] surveys but defer
detailed forecasts to future work. As we discussed previously
\citep{2018ApJ...867...26B}, the 21~cm sensitivity requirement assumed here
seems plausible considering HERA-350 will {\em image} some large scale modes
\citep{DeBoer:2016tnn} --- although the white noise approximation is rather
crude and should be refined in future work.

We caution that the strength of the [\CII] signal at the redshifts of interest
is quite uncertain. A broad range of estimates appear in the current
literature, depending on assumptions about: the correlation between [\CII]
luminosity and SFR at high redshift, the total star-formation rate density
(estimates from UV luminosity functions are sensitive to whether and how one
extrapolates to faint luminosities beyond current detection limits), and the
host-halo masses of [\CII] emitters. For example, our model values for
$\avg{I_{\CII}}$ are similar to a number of recent forecasts
\citep{2018arXiv180204804D, 2015ApJ...806..209S}, but are more than an order
of magnitude larger than some more pessimistic estimates in
\citet{2015ApJ...806..209S,2018arXiv181208135C}. In any case, at fixed
luminosity-weighted bias, the required noise scales quadratically with the
average specific intensity and so the reader can rescale our results according
to their preferred specific intensity model. For instance, in the case of
$\avg{I_{\CII}} = 20\,\text{Jy}/\text{sr}$ \citep{2018arXiv181208135C}, one
would require that $N_{\CII} = 2.6 \times 10^6$, $4 \times 10^6$, $6.2 \times
10^6\,(\text{Jy}/\text{sr})^2\,\Mpc^3$ for the L, M, and H models of the
[\CII] line, respectively at $z=8.34$. On the other hand, a more moderate
estimate of $\avg{I_{\CIIm}} = 100 \,\text{Jy}/\text{sr}$ \citep{
2015ApJ...806..209S,2018arXiv180204804D}, requires $N_{\CII} = 6.4
\times 10^7$, $1 \times 10^8$, $1.6 \times
10^8\,(\text{Jy}/\text{sr})^2\,\Mpc^3$ for the L, M, and H models of the
[\CII] line, respectively at $z=8.34$.

\begin{deluxetable}{cCCC}
\tablecaption{The noise power-spectrum for upcoming [\CII] surveys at $z=7.4$.
\label{tab:noise}}
\tablehead{\colhead{survey} & \colhead{$A_{\text{survey}}$} & \colhead{$A_{\text{pix}}$} & \colhead{$N_{\CIIm}$} \\ 
\colhead{} & \colhead{$(\text{deg}^2)$} & \colhead{$(\text{deg}^2)$} & \colhead{$((\text{Jy}/\text{sr})^2\,\Mpc^3)$} } 
\startdata
CCAT-p & 2 & 2.5\times10^{-4} & 2.66\times10^{9} \\
CONCERTO & 1.4 & 6.7\times10^{-5} & 2.04 \times 10^{9} \\
TIME & 1.3 \times 0.0084 & 6.7\times10^{-5} & 1.04\times10^9 \\
\enddata
\tablerefs{See \citet{2018arXiv181208135C} for more details.}
\end{deluxetable}

With the assumed noise and survey requirements for our fiducial model, we show
the resulting error bars in the {\em Upper} and {\em Middle} panels of
Figure~\ref{fig:b21_vs_k} for our particular choice of binning. At least for
the hypothetical surveys considered here, the 21~cm bias factor may be
recovered with good statistical precision. In other words, if sample-variance
limited sensitivity may be reached at $k=0.1\,\Mpc^{-1}$ in each line over a
common survey area of $\sim 100\,\text{deg}^2$, then a strong detection
appears feasible. Of course we have neglected sample variance contributions
from residual foregrounds among other complications, and so this should be
interpreted as a best-case scenario. On the other hand, increasing the common
survey area above $100\,\text{deg}^2$, for example, could help shrink the
error bars.

While our fiducial [\CII] survey is somewhat futuristic, we can also consider
the prospects with current, shortly upcoming surveys, specifically
CCAT-prime\footnote{\url{http://www.ccatobservatory.org}}
\citep{2018SPIE10700E..1MS},
CONCERTO\footnote{\url{https://people.lam.fr/lagache.guilaine/CONCERTO.html}}
\citep{Lagache:2018hmk}, and
TIME\footnote{\url{https://cosmology.caltech.edu/projects/TIME}}
\citep{Crites14}. We use the pixel noise values, $\sigma_{\text{pix}}
t_{\text{pix}}^{-1/2}$, for each survey from \citet{2018arXiv181208135C}. We
report the noise power spectrum at $z=7.4$ (assuming a pure white-noise
spectrum) in Table~\ref{tab:noise}. We generically find that
$N\sim2\times10^9\, (\text{Jy}/\text{sr})^2\,\Mpc^3$. If we assume a model
with $\avg{I_{\CIIm}}\sim500\,\text{Jy}/\text{sr}$ then even the
first-generation surveys reach our requisite noise. However, deeper surveys
will be needed in the case of the more pessimistic estimates of
$\avg{I_{\CIIm}}\sim100$ or $\sim20\,\text{Jy}/\text{sr}$. That being said,
our fiducial calculations also assume a larger survey area of
$100\,\text{deg}^2$. At $z=8.34$ we find a $\text{S}/\text{N}$ of $3.3$,
$2.7$, and $2.9$ for the L-M, M-H, and H-L models, respectively at
$k=0.1\,\Mpc^{-1}$ and bin width of $\Delta k=0.03\,\Mpc^{-1}$. Since the
number of modes scale with the square root of the survey area, we estimate
that CCAT-p might be able to recover a $\text{S}/\text{N}$ of $0.5$, $0.4$,
and $0.4$ for the L-M, M-H, and H-L models, respectively at
$k=0.1\,\Mpc^{-1}$. Including some higher $k$-modes, even this
first-generation survey might be capable of a marginal detection (if [\OIII]
can be surveyed as well), but this is only for our optimistic signal strength
model.

Since the strength of the [\CII] signal is likely a strong function of
redshift, the survey requirements should be less stringent at $z \sim 7$ than
the $z \sim 8$ case considered above. The main effect here should be from
redshift evolution in the average specific intensity; again, the noise
requirements scale with the average intensity squared. The required noise can
therefore be adjusted according to one's preferred model for redshift
evolution in the signal strength.

\section{Conclusions}\label{sec:conclusions}

We have shown that the amplitude of large-scale 21 cm fluctuations may be
inferred from measuring cross-power spectra between the 21 cm fluctuations and
each of two separate line-intensity maps, such as [\CII] or [\OIII]. Although
it has long been recognized that the cross-power spectrum between two fields
is more robust to foreground contamination than the auto-power spectrum of
either field alone, the amplitude of a single cross-power spectrum provides
only a product of two bias factors. We found that using a suitable combination
of three cross-power spectra
(Equations~\ref{eq:threefields}~and~\ref{eq:threefields_specific}) one can
instead infer the 21~cm bias alone to high accuracy.

Quantitatively, in the reionization model we considered, the accuracy reaches
percent-level on large scales ($k \sim 0.1-0.3\,\Mpc^{-1}$) during much of the
EoR. The inferred bias factor evolution can then be compared to that extracted
from the 21~cm auto spectrum. In principle, checking whether the 21~cm
auto-power spectrum follows linear-biasing on large scales might itself be a
good systematics check. However, linear biasing holds only over a limited span
of wavenumbers and early measurements may probe a small dynamic range in
spatial scale. Hence we believe that our three cross-spectra approach might
play an important role in confirming initial detections. Since our method
underestimates $P_{21,21}$ on intermediate scales, it can place informative
lower limits (i.e. $\sim 50\%$ of the true value) down to $k\sim1\,\Mpc^{-1}$,
depending on the stage of reionization. More work is necessary, however, to
see if there are some allowed reionization and line-intensity models where our
technique actually overestimates $P_{21,21}$.

Although we focused here on the case of 21~cm fluctuations during the EoR, the
method has broader applicability. For example, one can also estimate the bias
of the [\CII] and [\OIII] fluctuations by using a similar ratio of
cross-spectra. This should help circumvent the line-interloper problem that
presents a challenge for such surveys \citep[e.g.][]{kovetz2017:im_review}.
Since the ionized bubbles lead to scale-dependent biasing in the 21~cm field
on large spatial scales, the 21~cm case is an especially demanding
application, and we expect even better performance for [\CII], [\OIII], and
related lines.

In order to implement the strategy proposed here, there must be a coordinated
effort to probe the same regions on the sky over common redshifts in multiple
lines of interest. Ultimately, we envision line-intensity mapping surveys in
$N$ different lines, all probing the same cosmological volume. Among other
benefits, this will provide $N(N-1)/2$ measurements of the bias factor in each
line using the same basic technique outlined here.

\acknowledgments
We thank the anonymous referee for providing helpful comments. We thank Matt
McQuinn for the simulations used in this analysis. A.B. would like to thank
Todd Phillips for helpful discussions. A.B. was supported in part by the Roy
\& Diana Vagelos Program in the Molecular Life Sciences and the Roy \& Diana
Vagelos Challenge Award. The work of A.B. and F.V.-N. is supported by the
Simons Foundation.

\software{\texttt{colossus} \citep{2018ApJS..239...35D}, \texttt{matplotlib}
\citep{Hunter:2007}, \texttt{numpy} \citep{numpy:2011}, and \texttt{scipy}
\citep{scipy:2001}.}

\bibliography{references}

\end{document}